\documentclass[aps,superscriptaddress,notitlepage]{revtex4-1}
\usepackage{graphicx}
\usepackage{amsmath,amssymb,amsfonts}
\newcommand{\vrr}{\mathbf{r}}
\newcommand{\vn}{\mathbf{n}}
\newcommand{\vk}{\mathbf{k}}
\newcommand{\vg}{\mathbf{g}}

\newcommand{\vh}{\mathbf{h}}
\newcommand{\vj}{\mathbf{j}}

\begin{document}
\title{A hydrodynamic approach to electron beam imaging using a Bloch wave representation}
\author{S. Rudinsky}
\affiliation{Department of Mining and Materials Engineering\\ McGill University\\ 3610 University, Montreal, Canada, H2T 2X1}
\author{R. Gauvin}
%\author[1]{R. Gauvin}
\affiliation{Department of Mining and Materials Engineering\\ McGill University\\ 3610 University, Montreal, Canada, H2T 2X1}
 \begin{abstract}
  Calculations of propagating quantum trajectories associated to a wave function provide new insight
into quantum processes such as particle scattering and diffraction. Here, hydrodynamic calculations
of electron beam imaging under conditions comparable to those of a scanning or transmission electron
microscope display the mechanisms behind different commonly investigated diffraction conditions. The
Bloch wave method is used to propagate the electron wave function and associated trajectories are computed to map the wave function as it propagates through the material. Simulations of the two-beam
condition and the systematic row are performed and electron diffraction is analysed through a real space
interpretation of the wave function. In future work, this method can be further coupled with Monte Carlo
modelling in order to create all encompassing simulations of electron imaging.
 \end{abstract}
\maketitle
\section{Introduction}
The wave-particle duality of electrons causes a segregation between the types of simulation
techniques employed in the field of electron microscopy. Simulations are categorized as
either image simulations, where the probability distribution of the electron
wave function is used to obtain contour maps relating to the exit plane or
diffraction pattern, or particle scattering simulations using classical methods
to obtain intensities associated mostly to X-ray emission events and particle
penetration \cite{kirkland2010advanced}. Image simulations use techniques such
as multislice \cite{Allen201511,VANDENBROEK201589,DWYER2010195} or Bloch wave
\cite{Pennycook1990PhysRevLett,Winkelmann2007414,FINDLAY200365,ROSSOUW2003299}
which simulate the probability density in real or reciprocal space of the
electron upon exit of the material. Conversely, scattered particle trajectories
are typically simulated through Monte Carlo techniques where electrons are
assumed to be classical spheres that undergo a forward scattering random walk
process where the scattering and energy loss parameters are calculated by
physical models \cite{gauvin2006win,salvat2015penelope}. As of yet, there is no
technique which simultaneously simulates both the wave and particle
characteristics of electrons within the confines of an electron microscope.  \\

Here, we couple the Bloch wave representation of the electron wave function with the
propagation of quantum trajectories to simulate electron-matter interactions
inside a crystalline material under various probing conditions. The quantum
trajectory method arises from the hydrodynamic formulation of a quantum process
\cite{trahan2006quantum,PhysRevLett.98.066803}. Given an initial position, the particle will follow a
specific path dictated by the wave function. The uncertainty then comes in the
choice of the initial position, preserving the non-locality of the method \cite{sanz2012trajectory}. Such
simulations have mostly been performed for particle diffraction experiments and
small scale quantum processes \cite{sanz2010understanding,Sanz2002Diff,EFTHYMIOPOULOS2012438}. In
previous work, the method was applied in 2D to simulate the time-dependent
propagation of a Gaussian wave-packet under conditions similar to those of an
SEM \cite{RudinskyNearfield2016}. It was found however that there were a number
of limiting numerical factors, such as the energy bandwidth, grid size, and
film thickness, which restricted the applicability of a time-dependent
propagation scheme \cite{RudinskyNearfield2016}. The algorithm developed in this study constitutes a completely new and different approach. The trajectories are no longer calculated using a spectral decomposition and the split-operator method is substituted for the Bloch wave method. Other work has been done using
a multislice approach to simulate a scanning transmission electron microscope
(STEM) probe located at different positions along a unit cell
\cite{Zhang2015JMicro,zeng2011simulation}. However, there was little analysis
done of the beam interaction with the material, specifically at different
probing conditions \cite{Zhang2015JMicro}. The use of the multislice method
also limits usability of the calculation because the trajectories may only be
computed within the confines of the chosen grid. An important advantage of the
Bloch wave method is the possibility of computing the wave function at any
point in space without being restricted to a structured grid. Furthermore,
computations may be performed at lower accelerating voltages, making the method
applicable for wave function simulations at energies typically used in scanning
electron microscopes (SEM) \cite{Winkelmann2007414,PICARD201471}. With this, Bloch wave
calculations coupled with quantum trajectories have been previously investigated by Cheng
\textit{et al.} \cite{Cheng_2018}. While they displayed computations at normal
incidence and simulations of electron backscattered diffraction images (EBSD),
criteria such as the number of beams used and the initial wave function of the
EBSD computation method were not indicated, making it difficult to reproduce
their findings. There was also no explanation of the calculation process
used in the EBSD simulations and consequently other, more in depth and reproducible studies, are
necessary. \\

In this study, the electron wave function along the particle path was computed using the Bloch
wave expression to simulate electrons travelling through a single crystal of
Cu. Simulations were performed at 200 and 30 keV to distinguish the difference
in electron transport in TEM versus SEM-like conditions. The initial positions of the
trajectories were chosen uniformly to map out the wave function over the entire
unit cell. A variety of probing conditions were also investigated, such as the
(100) zone axis, the two beam Bragg condition, and the systematic row
condition. Trajectory simulations in these conditions provide information to
the origins of various contrasts typically observed. The quantum potential and
quantum force are also computed at the exit plane. These values represent the
quantum effects which cannot be explained through classical particle
propagation \cite{trahan2006quantum}. As a result, these parameters show how
electron propagation through a crystalline material leads to diffraction
phenomena. It is shown through the quantum force that at normal incidence,
electrons are drawn to and from the atom columns, resulting in their
channelling through the material. This is also seen in the systematic row case
where particles between atom columns are drawn towards them causing variations
in the edges of the band contrasts.

%%%%%%%%%%%%%%%%%%%%%%%%%%%%%%%

\section{Method} 
\begin{figure*}
 \centering
 \includegraphics[width=\textwidth]{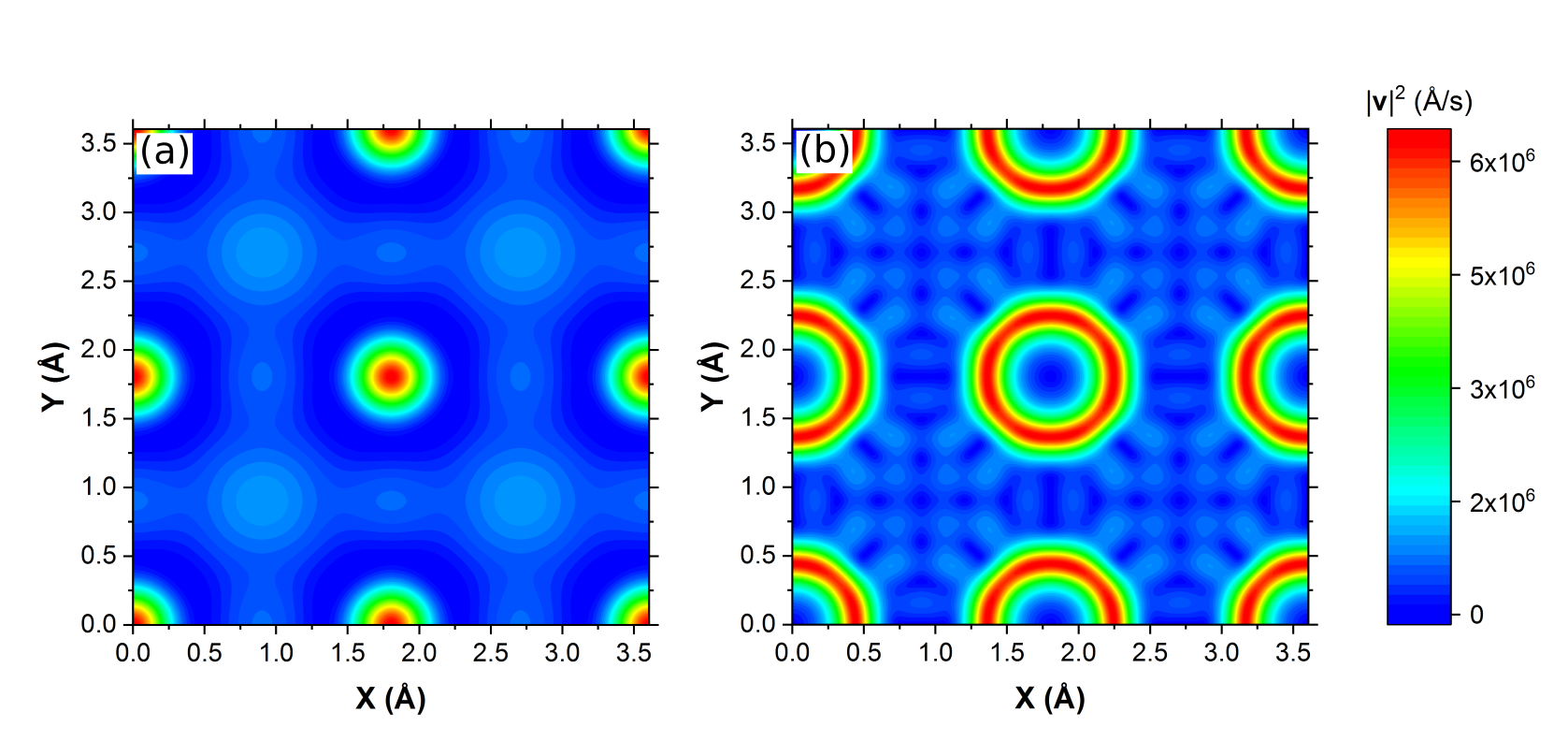}
 \caption{Exit wave function of Cu(100) zone axis orientation at 200keV and a thickness of 500 \AA.}
 \label{fig:200keV-ZA-EW}
\end{figure*}

A Bloch wave expression was used to compute the electron wave
function along the trajectory paths. The wave function, $\Psi(\vrr)$, of an
electron in a periodic potential can be expressed as a sum of Bloch waves
weighted by excitation coefficients $\alpha_\vg^{(j)}$ for each Bloch wave $j$
\cite{de2003introduction},
\begin{equation}
 \Psi(\vrr) = \sum_\vg\left(\sum_j\alpha_\vg^{(j)}C_\vg^{(j)}e^{2\pi i\gamma^{(j)}z}\right)e^{2\pi i(\vk_0+\vg)\cdot\vrr}
 \label{eq:BWdecomp}
\end{equation}
The coefficients $C_\vg^{(j)}$ and contributions $\gamma^{(j)}$ of each Bloch wave are obtained by solving the eigenvalue equation,
\begin{equation}
 \vk_0 - (\vk^{(j)}-\vg) C^{(j)}_{\vg}+\sum_{\vh\neq\vg}U_{\vg-\vh}C^{(j)}_{\vh}=2\gamma^{(j)}k_0C^{(j)}_{\vg}
 \label{eq:EigEqFinal}
\end{equation}
where $C_\vg^{(j)}$ are the eigenvectors and $\gamma_\vg^{(j)}$ are the
eigenvalues \cite{de2003introduction}. The factors $U_\vg$ are obtained from
the Fourier coefficients, $V_\vg$, of the electrostatic potential,
\begin{equation}
 U_\vg = \frac{2m_0e}{h^2}V_\vg
 \label{eq:U-from-V}
\end{equation}
for a particle of mass $m_0$, charge $e$ and where $h$ is Plank's constant. The
Fourier coefficients, $V_\vg$, were computed using a summation over the
pairwise contributions of the $N$ atoms in the unit cell as, 
\begin{equation}
 V_\vg = \sum_{\vn\in N}f_\vg e^{-2\pi i(\vg\cdot\vn)}
 \label{eq:vg}
\end{equation}
 where $f_\vg$ is a parametrization of scattering factors tabulated by Kirkland \cite{kirkland2010advanced}.
 \begin{equation}
  f_\vg = \sum_{k=1}^3 \frac{A_k}{|\vg|^2+B_k}+C_ke^{-D_k|\vg|^2}
  \label{eq:fg}
 \end{equation} 
 This is in contrast to the parametrization of the real space potential utilised in \cite{RudinskyNearfield2016}. Here, only the Fourier coefficients of the potential are required. Absorption was not included in the simulations performed in
 this study, which is why the imaginary term of the electrostatic potential is
 neglected in the above derivations. Simulations were performed for copper and
 the values of the parametrization factors in Eq. \ref{eq:fg} are displayed in
 Table \ref{tab:param}
\begin{table}
 \caption{Parametrization factors for Cu \cite{kirkland2010advanced}}
 \centering
 \begin{tabular}{cccc}
  \hline\hline
  $A_k$ (\AA$^{-1}$)&0.358774531 &1.76181348 &0.636905053 \\
  $B_k$ (\AA$^{-2}$)& 0.106153463&1.01640995 &15.3659093 \\
  $C_k$ (\AA)& 0.00744930667&0.189002347 &0.229619589 \\
  $D_k$ (\AA$^{2}$)& 0.0385345989&0.398427790 &0.901419843 \\
  \hline
 \end{tabular}
\label{tab:param}
\end{table}
After the beams with zero structure factor were eliminated, Bethe potentials
were used to further limit the number of beams required for the computation.
Beams are separated into weak and strong beams depending on the following
criteria \cite{ZUO1995375,WANG201635,PICARD201471}.  
\begin{align}
 &\frac{|s_\vg|}{\lambda U_\vg}\leq c_s \mbox{\hspace{0.5cm}(strong)} \\
 c_s < &\frac{|s_\vg|}{\lambda U_\vg} \leq c_w \mbox{\hspace{0.5cm}(weak)}
\label{eq:beamselection}
 \end{align}
where $s_\vg$ is the excitation error and $\lambda$ is the wave length. Beams
that are considered strong are used for the diagonalization, while those that
are weak contribute as perturbation factors to the entries of the dynamical
matrix. The values $(c_s,c_w)$ were chosen to ensure the difference between the
full dynamical matrix and the reduced matrix using Bethe potentials was within
the order of $10^{-3}$. The difference between the intensities is defined as
follows \cite{WANG201635}, 
\begin{equation}
 \delta I = \sqrt{\frac{1}{N_s}\frac{1}{T}\sum_{N_s}(I_\vg^f-I_\vg^s)^2}
 \label{eq:dI}
\end{equation}
where $N_s$ is the total number of strong beams, $I_\vg$ is the intensity of
beam $\vg$ and $T$ is the maximum thickness used in the calculation. The order
of magnitude chosen was done to limit oversampling of the electron wave
function. Quantum trajectories are very sensitive to small local perturbations
and therefore, in order to simultaneously ensure smoothness of the trajectories
and exactness of the wave function, an intensity difference on the order of
$10^{-3}$ was chosen to be a reasonable limit. Once the real space wave
function is computed, it can be used to calculate trajectories of the
associated quantum particle. The quantum trajectory method, described in
previous work \cite{RudinskyNearfield2016}, provides a visual representation of
electron-matter interactions as the beam is propagated through a material.
In this formulation, the polar form of the wave function is used to solve the
Schr\"{o}dinger equation resulting in a continuity equation and a quantum form
of the Hamilton-Jacobi equation (QHJ) \cite{trahan2006quantum}. The QHJ differs
from its classical counterpart by an additional term called the quantum
potential, $Q$, which accounts for all effects arising from the quantum nature
of the system. The quantum potential is expressed as follows, 
\begin{equation}
 Q = -\frac{\hbar^2}{2m}\frac{1}{R} \nabla^2R
 \label{eq:Qp}
\end{equation}
where $R = \sqrt{|\Psi|^2}$ \cite{trahan2006quantum}. From this, the quantum
force, $f_q$, exerted on the particle may also be obtained since $f_q = -\nabla
Q$. From a given initial position, $\vrr_0$, a quantum trajectories position at
a further point in time is computed by integrating over the velocity field,
$\dot{\vrr}$ \cite{sanz2013trajectory}, 
\begin{equation}
 \vrr(t) = \vrr_0(t_0) + \int_{t_0}^t \dot{\vrr} dt
 \label{eq:propagate}
\end{equation}
From the definition of the flux, $\vj$, for the probability density and the continuity equation \cite{trahan2006quantum},
\begin{equation}
 \dot{\vrr} = \frac{\vj}{|\Psi(\vrr)|^2} = \frac{\hbar}{m}\Im\left(\frac{\nabla\Psi(\vrr)}{\Psi(\vrr)}\right)
 \label{eq:vfield}
\end{equation}
Because $\Psi(\vrr)$ is expressed as a sum of Bloch waves, its gradient has an
analytical solution improving the speed of computing Eq. \ref{eq:vfield}.
With the velocity field at a specific point along the trajectory, Eq.
\ref{eq:propagate} is solved using a second order Runge-Kutta. Since the Bloch
wave method generates a time-independent solution to the Schr\"{o}dinger
equation, Eq. \ref{eq:propagate} is in fact solved for increments of thickness,
generating time-independent trajectories, i.e. a continuous flow. The initial
position of each trajectory was chosen systematically across the entrance
surface of the unit cell. This was done to map out the progression of all
portions of the wave function as it travels through the crystal. Trajectories
were either positioned on a $10\times10$ grid covering a single unit cell, or 50
in a line parallel to the $x$-axis so that they may be viewed in a 2D
projection. An example of the exit wave calculated using the Bloch wave method
and its associated velocity field for Cu(100) in the zone axis orientation is
displayed in Figure \ref{fig:200keV-ZA-EW}. Here, the magnitude of the velocity
field is shown.  
Simulations were performed at 200 keV and 30 keV. A variety of conditions were
considered. Simulations were performed at normal incidence in the zone axis
orientation of Cu(100), in the two beam condition for the 200 reflection, and in the case of a
systematic row. Trajectories were propagated to a maximum thickness of 500 \AA.
In all cases, the incident wave function was chosen to be a plane wave.
\section{Results and discussion}
\subsection{Normal incidence}
\begin{figure*}
\centering
 \includegraphics[width = \textwidth]{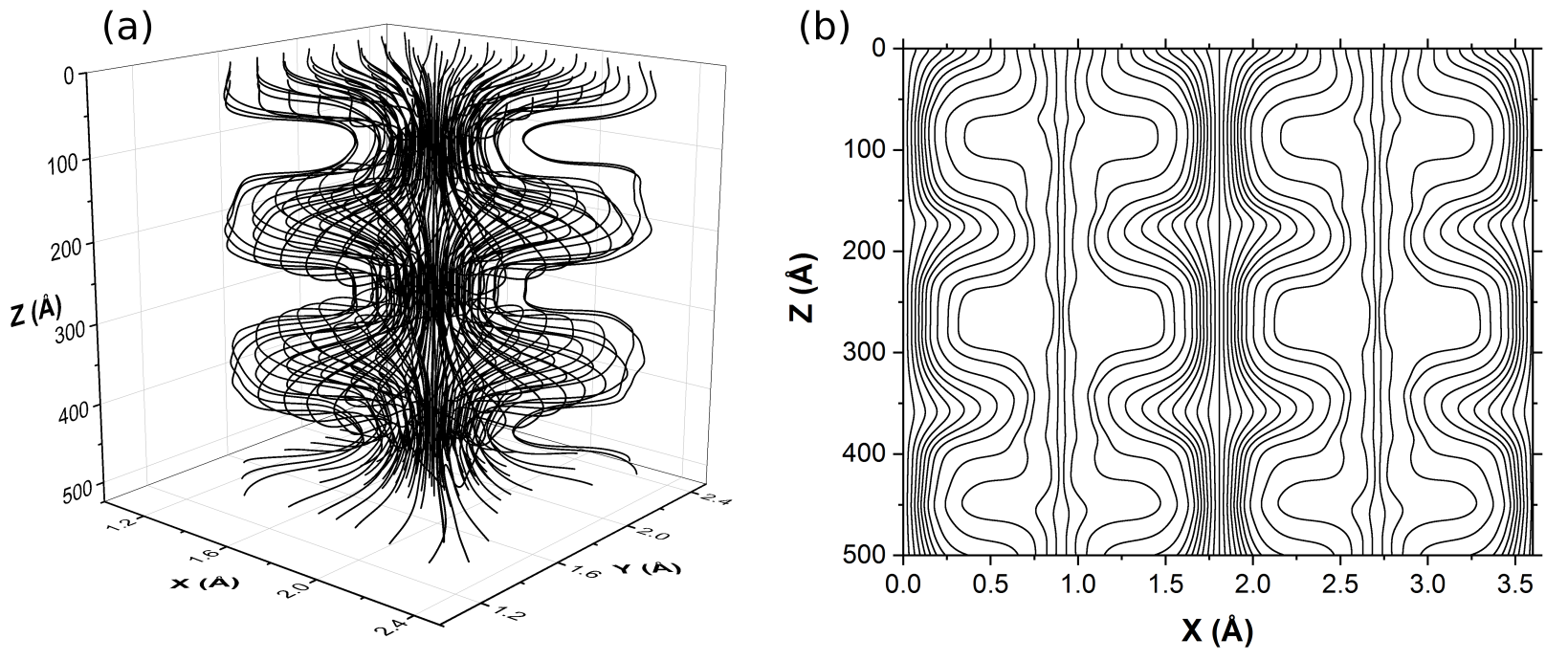}
 \caption{(a) 3D quantum trajectories at 200 keV of Cu(100) in zone axis
 orientation and (b) 2D projection on the $x-z$ plane of trajectories computed
 under the same imaging conditions but with 50 trajectories placed along a line.} \label{fig:200keV-ZA-traj}
\end{figure*}

\begin{figure*}
\centering
\includegraphics[width = 0.75\textwidth]{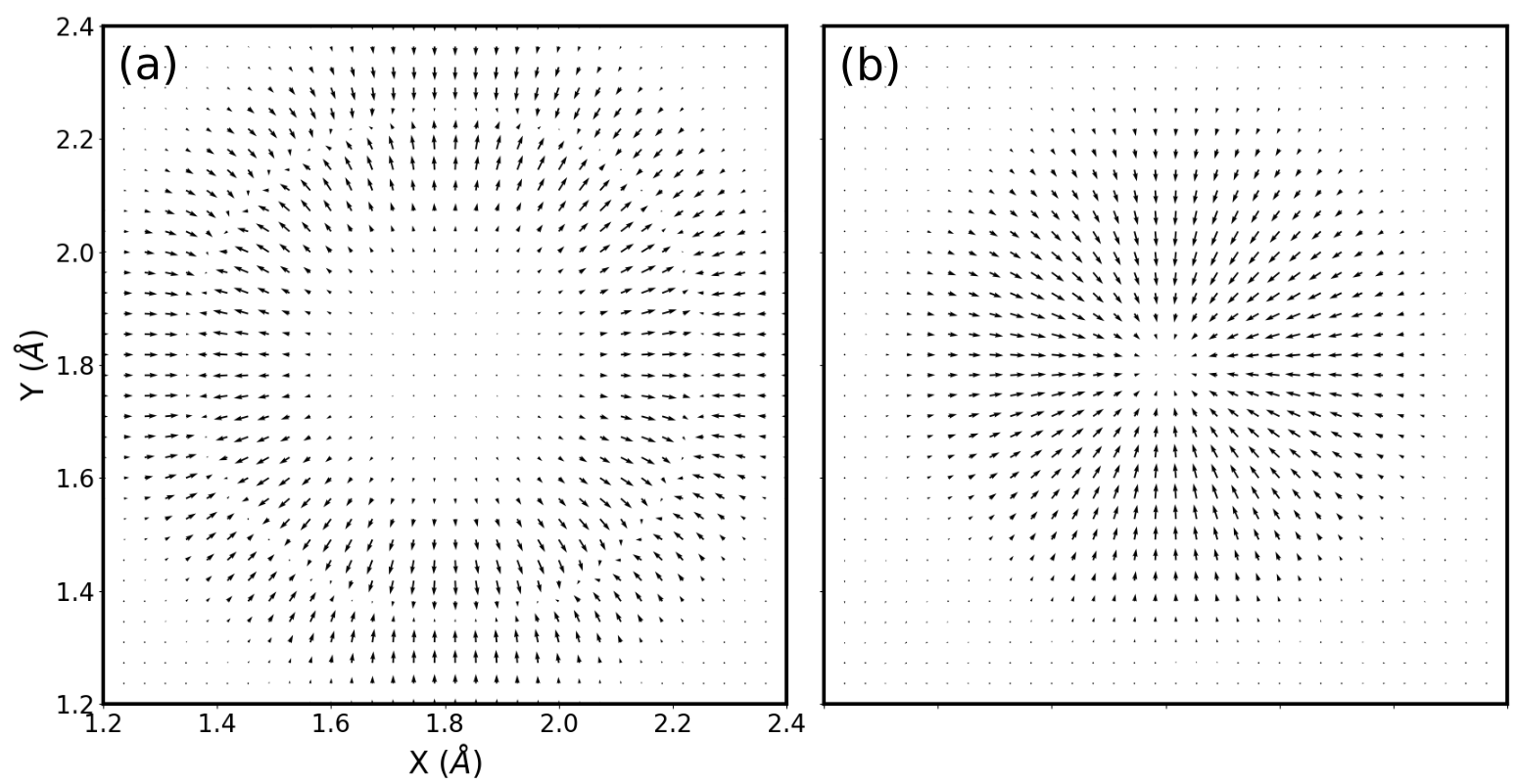}
\caption{(a) Quantum force generated by interaction between electron wave
function and material around a single atom column and (b) electrostatic force
of Cu.} \label{fig:200keV-ZA-force}
\end{figure*}

Simulations were first performed at normal incidence in the (100) zone axis
orientation for copper. For an incident energy of 200 keV, $c_s=80$ and $c_w=90$ which resulted in a
computation of 29 strong beams and 8 weak beams. Trajectories were first
positioned equidistant from each other in a $10\times10$ grid for an entire 3D
analysis of the wave function progression. Then, to generate a 2D projection of
the system, 50 trajectories along a line parallel to the $x$-axis in the center
of the unit cell were initiated. Figure \ref{fig:200keV-ZA-traj} (a) contains
the 3D depiction of quantum trajectories passing through a 500 \AA\ thick Cu
crystal at 200 keV around the center atom column of the unit cell. Figure
\ref{fig:200keV-ZA-traj} (b) is a 2D projection on the $x-z$ plane under the
same simulation conditions. Contrarily, for the afore mentioned figure, 50 trajectories were simulated along a line
parallel to the $x$-axis, cross-sectioning the center of the unit cell. As a
result, the behavior of these trajectories can be more clearly observed as they
propagate through the material. Since the image represents a 2D projection, the
patterns generated along the atom columns include oscillations coming out
of the plane.  
As the trajectories propagate through the material, they are consistently
attracted and repulsed by the atom columns. This effect is caused by the
quantum force. Figure \ref{fig:200keV-ZA-force} shows the vector fields of the
quantum and electrostatic forces acting around a single atom column.
If the only force acting on the propagating electrons was the electrostatic
force generated by the material, then the particles would be strongly attracted
towards the nucleus and would collapse without further transport. However, the
total force exerted on the particles is the sum of the electrostatic force and
the quantum force described by the hydrodynamic theory
\cite{chattaraj2010quantum}. The quantum force, as seen in Figure
\ref{fig:200keV-ZA-force} (a), contributes the repulsive force acting upon the
trajectories which pushes the electrons to channel between and through the atom
columns. The electrons are pulled towards the nucleus by both the quantum and
electrostatic forces until a critical radius where the repulsive quantum force
takes over and ensures that the particles do not collapse but continue to
propagate through. The electrostatic force is on the order of 500
N while that of the quantum force is 1600 N showing that the magnitude of the
quantum force supersedes that of the electrostatic, and this is evident by the
path the electron trajectories take. This also demonstrates the quantum force's
role in elastic scattering, where the magnitude and direction of a scattering
event can be described in terms of the quantum force exerted on the particle.
The explanation for these events through the hydrodynamic theory can be coupled
with the conventional dynamic theory to provide a more complete explanation of
electron-matter interactions during transmission in an electron microscope.  \\

\begin{figure*}
 \centering
 \includegraphics[width=\textwidth]{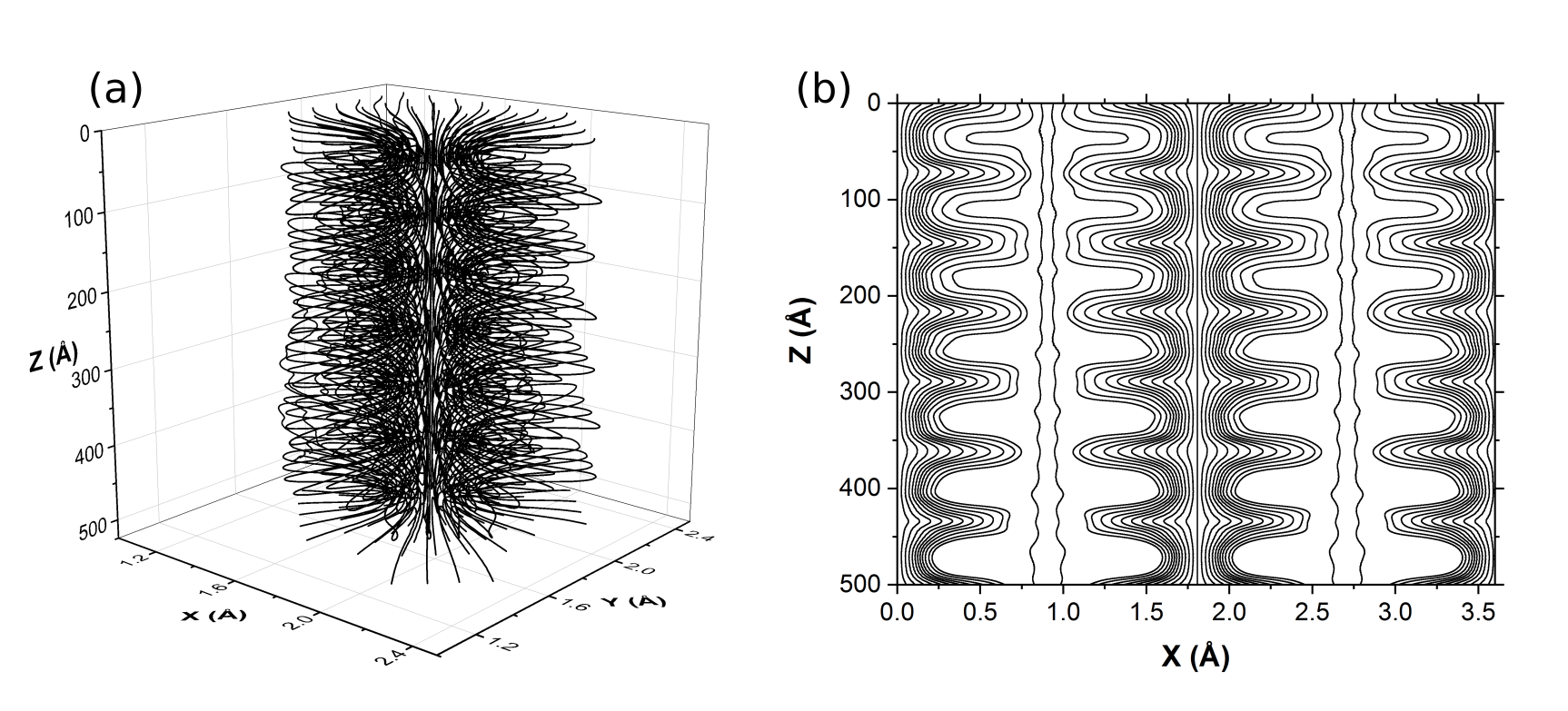}
 \caption{Simulations of a 30 keV electron beam incident on Cu(100) zone axis,
 (a) 3D representation of trajectories around a single atom column and (b) the
 2D projection with 50 trajectories positioned on a line parallel to the
 $x$-axis.} \label{fig:30keV-traj}
\end{figure*}

\begin{figure*}
 \centering
 \includegraphics[width=\textwidth]{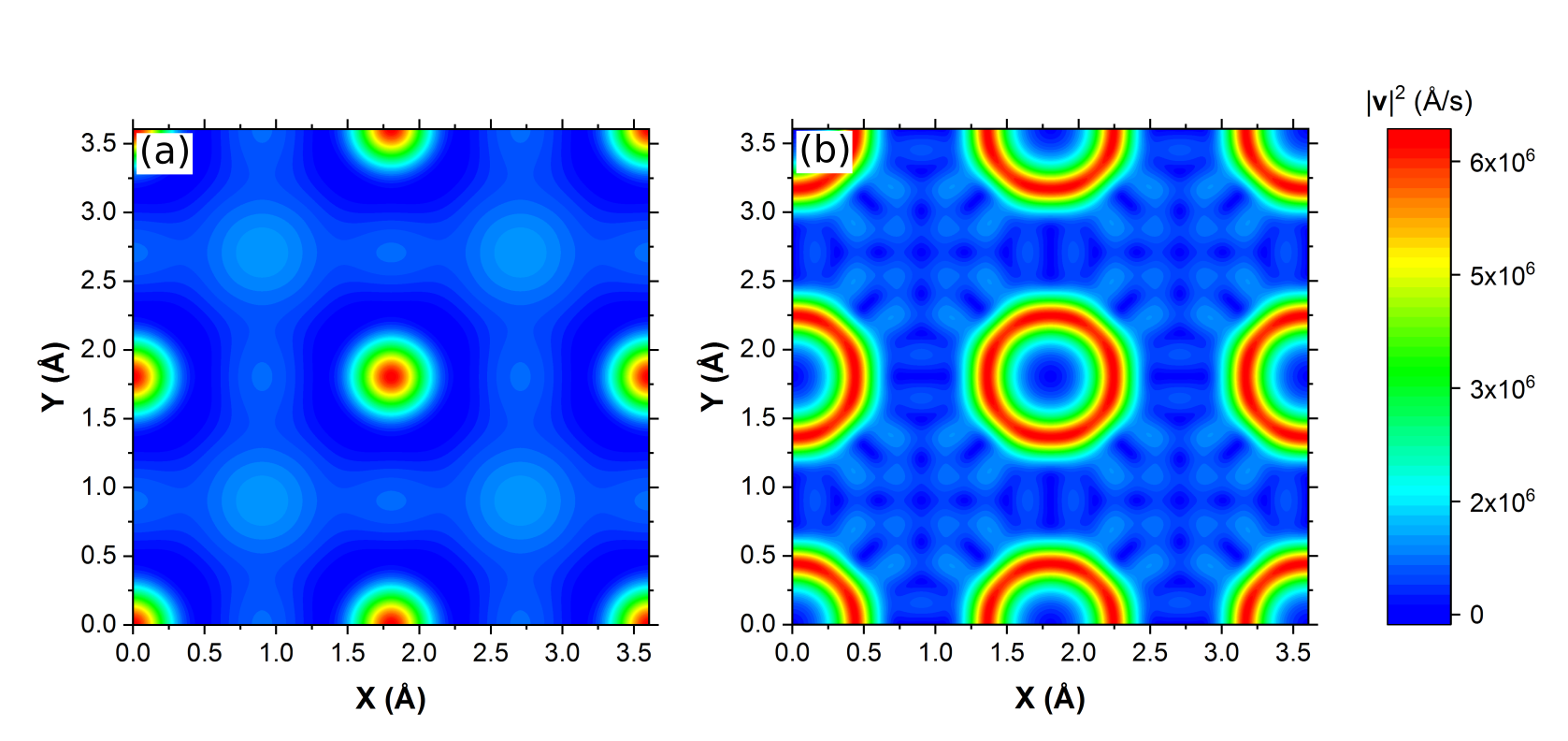}
 \caption{(a) Intensity of exit wave function and (b) associated velocity field
 with color scale for simulation done at 30 keV in Cu(100) zone axis
 orientation.} \label{fig:30keV-ZA-EWVF}
\end{figure*}

\begin{figure}
	\centering
	\includegraphics[width=0.5\textwidth]{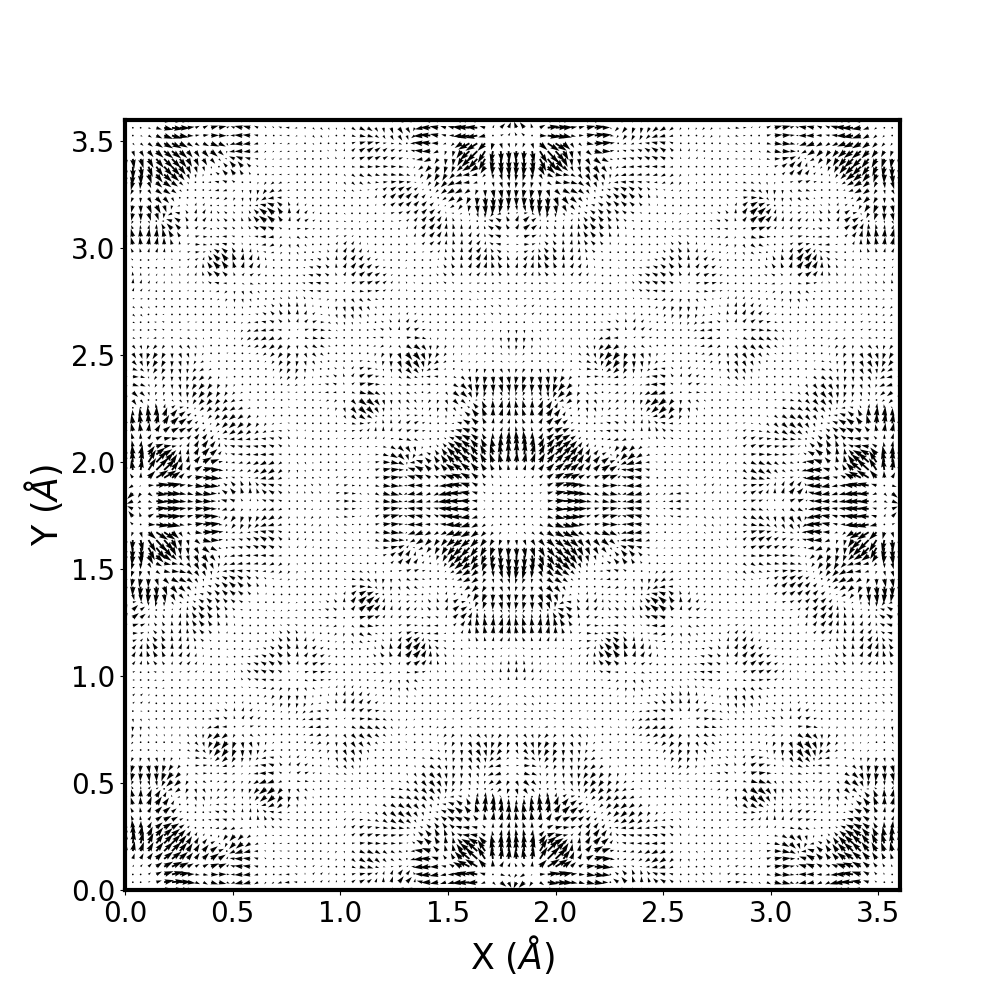}
	\caption{Quantum force at 30 keV across entire unit cell.}
	\label{fig:30kV-fq}
\end{figure}

Simulations were also performed at 30 keV to replicate wave function
propagation in the energy range of a scanning electron microscope (SEM). Here,
57 strong beams and 20 weak beams were used in the computation. Figure
\ref{fig:30keV-traj} shows the 3D representation of the trajectories around a
single atom column and the 2D projection across the entire unit cell.
Here, because the electron energy is significantly lower, there are many more
scattering events that will occur causing more modulations of the wave function
as it travels through the material. At a foil thickness of 500 \AA, 30 keV is
still large enough where the entire beam is transmitted through the film. The
wave function however undergoes significantly more coherent scattering which is
seen by the modulations in the trajectories. A consistency does get reached
after a few hundreds of angstrom where there is less dispersion between the
atom columns and the wave function becomes more confined to the sinusoidal
displacements and group channeling around the atom columns. The higher
frequency oscillations is again due to the action of the quantum force and this
is seen in the velocity field of the wave function. Figure
\ref{fig:30keV-ZA-EWVF} shows the magnitude of the velocity field across the
entire unit cell and its associated exit wave function.  
While the intensity of the exit wave function is greatest at the atom columns,
the velocity field shows that the particles' radial velocity increases near the
contours of the atom columns. The direction of the velocity is away from the
columns indicating again the channeling effect. The quantum force at 30
keV is, in contrast, much more chaotic due to high potential for interactions
and scattering at lower electron energies. Figure \ref{fig:30kV-fq} shows the quantum force
around an atom column at 30 keV. 

\subsection{Two beam condition}
\begin{figure*}
\centering
 \includegraphics[width = \textwidth]{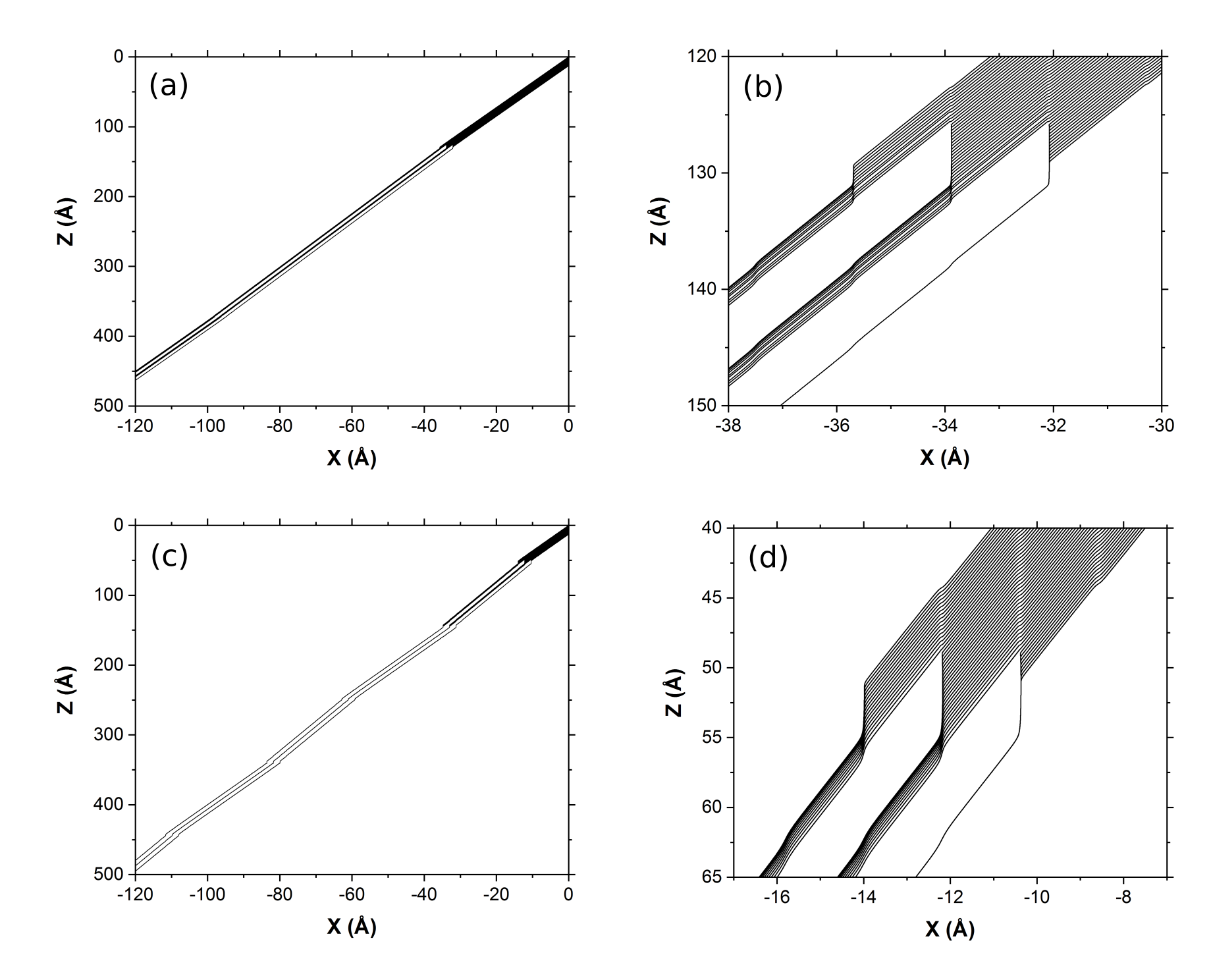}
 \caption{Quantum trajectories in the two beam condition for $\vg=(200)$ for Cu
 at electron energies of (a) 200 keV and (c) 30 keV. A zoom in of the full
 simulated trajectories at a thickness at which the trajectories are separated
 for (b) 200 keV and (d) 30 keV is also
 displayed.} \label{fig:Bragg-Traj}
\end{figure*}

\begin{figure*}
 \centering
 \includegraphics[width=\textwidth]{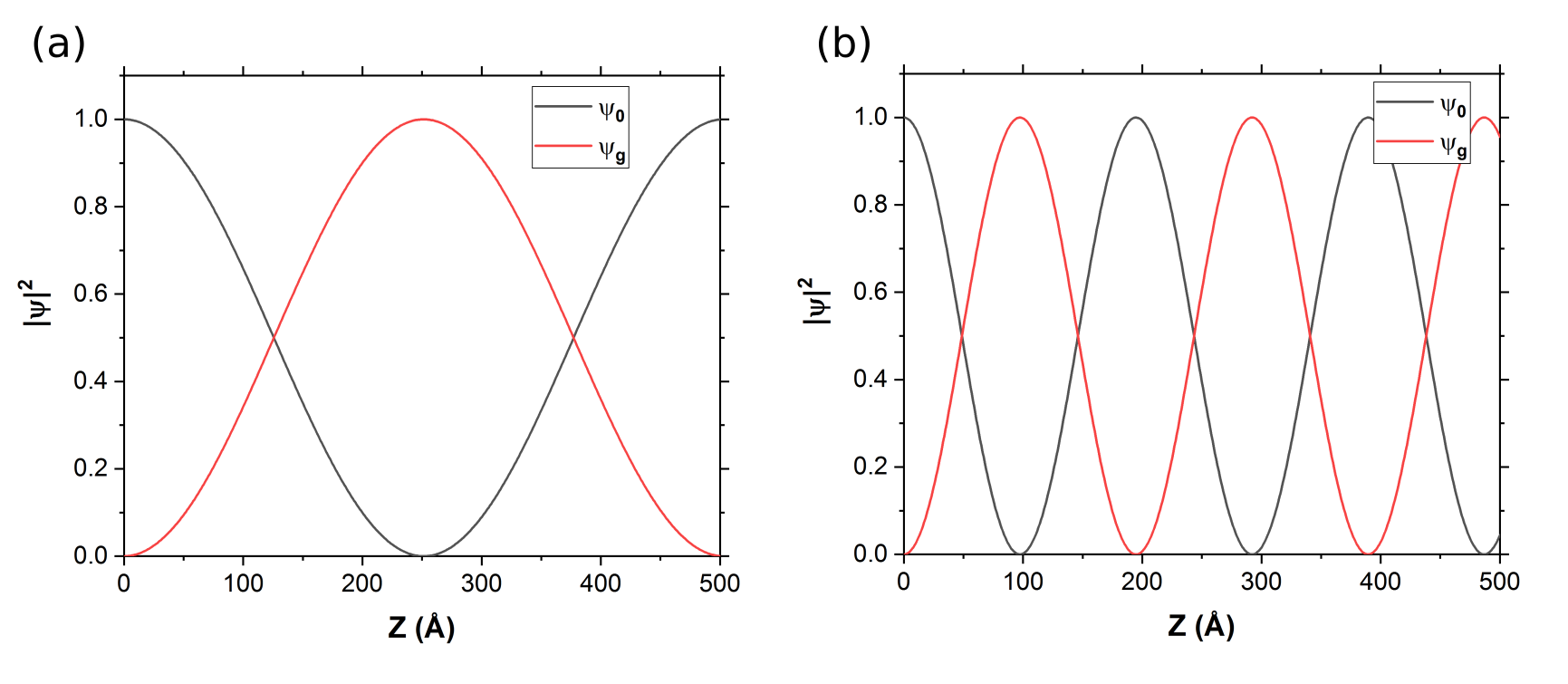}
 \caption{Beam intensity as a function of thickness for the $\vg_{200}$ two beam condition at electron energies of (a) 200 keV and (b) 30 keV.}
 \label{fig:psig}
\end{figure*}

\begin{figure*}
  \includegraphics[width=\textwidth]{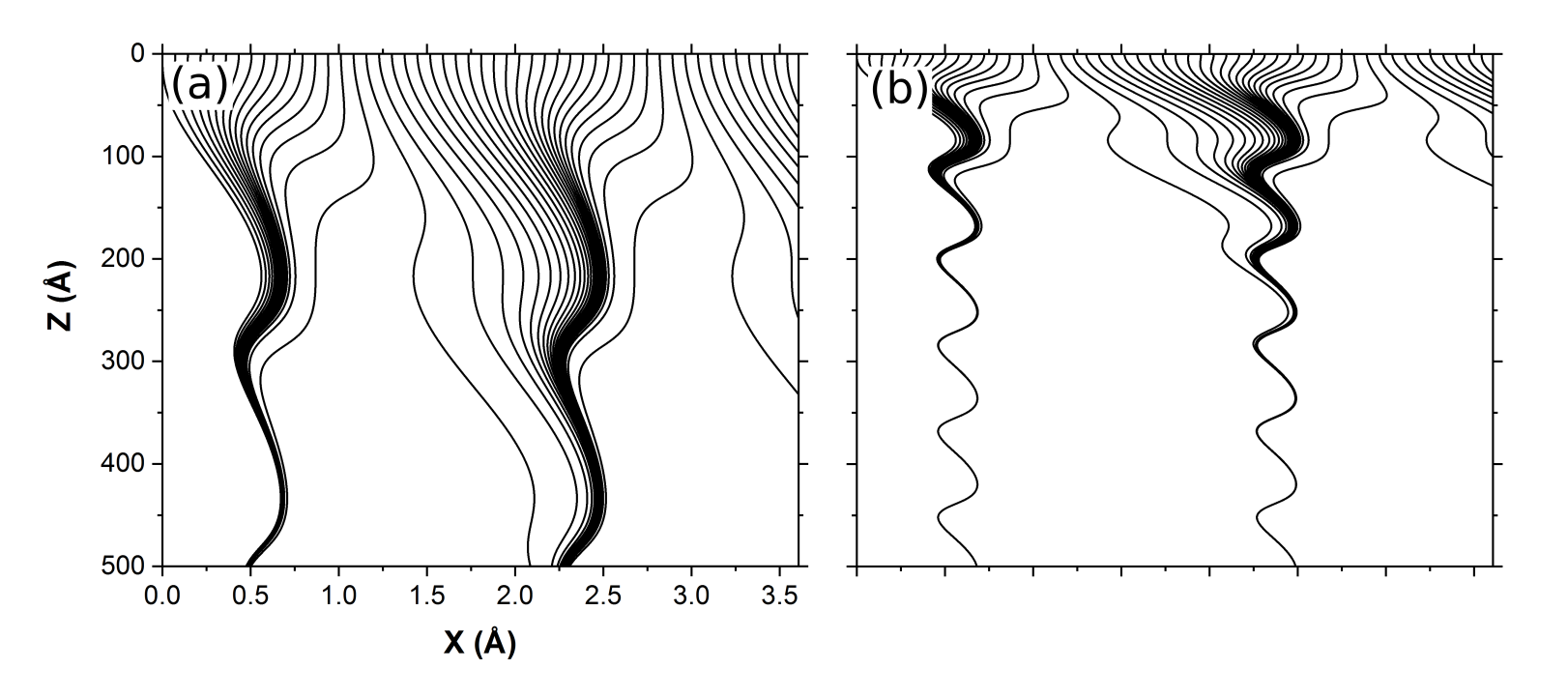}
  \caption{Quantum trajectories computed from two beams, $\psi_\mathbf{0}$ and $\psi_\vg$, where $\vg = (200)$. Propagation is at normal incidence and the electron energies are (a) 200 keV and (b) 30 keV respectively.} 
  \label{fig:TBNI}
 \end{figure*}

Trajectories in the two beam approximation were simulated to display the
particle paths under ideal diffraction conditions. Here, only two beams were
considered in the Bloch wave calculation, $\vg_{000}$ and $\vg_{200}$.
Simulations were again performed at 200 keV and 30 keV. As is
the convention, a transverse component whose
magnitude is equivalent to a tilt to the Bragg angle is added to the incident
wave vector, instead of tilting the specimen itself. The
transverse component added to the incident wave vector before normalization to
the electron wave length was $|\vk_t|=0.2766$ \AA$^{-1}$. This ensured that the exact
Bragg condition was met for $\vg_{200}$. The associated quantum trajectories are
displayed in Figure \ref{fig:Bragg-Traj}.  
In the two beam condition, only the primary and diffracted beam contribute to
the final wave function. Here, diffraction of the plane wave by the crystal
causes the trajectories to be separated into what are the primary and
diffracted beams. These trajectories then propagate accordingly resulting in
the pattern that would be detected at the diffraction plane. Figures
\ref{fig:Bragg-Traj}(a) and (c) display the entire propagation where once the
plane wave diffracts, the coupled groups of trajectories continue as such.
Figure \ref{fig:Bragg-Traj}(b) and (d) show a close up to the thickness where
the trajectories originally separate showing exactly how the wave function is
diffracted in the two beam condition. The small ripples in the grouped trajectories could be
effects of the "beating" caused by the two types of Bloch waves which have been
previously related to the appearance of thickness fringes \cite{egerton2013electron}. An important stipulation of the
theory behind the quantum trajectory method is that
the trajectories may not cross paths at a specified moment in time \cite{pladevall2012applied}. Therefore,
there cannot be crossing from one beam to another in configuration space.
However, the contributions of the two components of the travelling function do
oscillate accordingly. Figure \ref{fig:psig} shows the intensity of both waves
$\psi_\mathbf{0}$ and $\psi_\vg$ as a function of distance
for the electron energies considered here.  
In accordance with diffraction theory, because the incident wave
is tilted to exactly the Bragg condition and absorption is neglected, the intensities of each contributing
beam oscillate continuously and the full intensity is completely transferred
from one to the other \cite{cowley1995diffraction,reimer2008transmission}.
The extinction distance,
$\xi_\vg$, of a beam $\vg$ is computed as \cite{de2003introduction,wang1995elastic}, 
\begin{equation}
	\frac{1}{\xi_\vg} = \frac{|U_\vg|}{|\vk_0+\vg|\cos\alpha}
 \label{eq:xig}
\end{equation}
 where $\alpha$ is the angle between $\vk_0+\vg$ and the surface normal of the
 material. For Cu, computed using the scattering factors obtained by Kirkland
 \cite{kirkland2010advanced}, $\xi_\mathbf{200}$ is 431.1 and 166.9 \AA\ for
 200 keV and 30 keV respectively. The wave lengths of the sinusoidal functions
 governing the intensities of both beams in Figure \ref{fig:psig} correspond
 exactly to their respective extinction distances. This demonstrates the so
 called \textit{pendell\"{o}sung} \cite{doi:10.1080/01418617908234846}.  The
 total wave function in real space is however continuous which is portrayed
 through the propagation of trajectories. When the incident beam is tilted outside of the Bragg condition, the structure is lost and the trajectories behave in a different way. Figure \ref{fig:TBNI} shows trajectories computed from the same two beams but at normal incidence to the sample surface. 
Here, both the primary beam $\psi_\mathbf{0}$ and $\psi_\vg$ are the sole beams used in the computation. Because the Bragg condition is not satisfied, the beam is not diffracted and instead oscillations similar to those seen in the zone axis orientation are apparent. This indicates that outside the Bragg condition, if only two beams contribute to the Bloch wave expression, the wave function is only incoherently scattered by the atom columns. Evaluating the problem from a hydrodynamic approach can aid in explaining the two beam diffraction process in real space with the evolution of the wave function.  
 
 \subsection{Systematic row} 
\begin{figure*}
 \includegraphics[width=\textwidth]{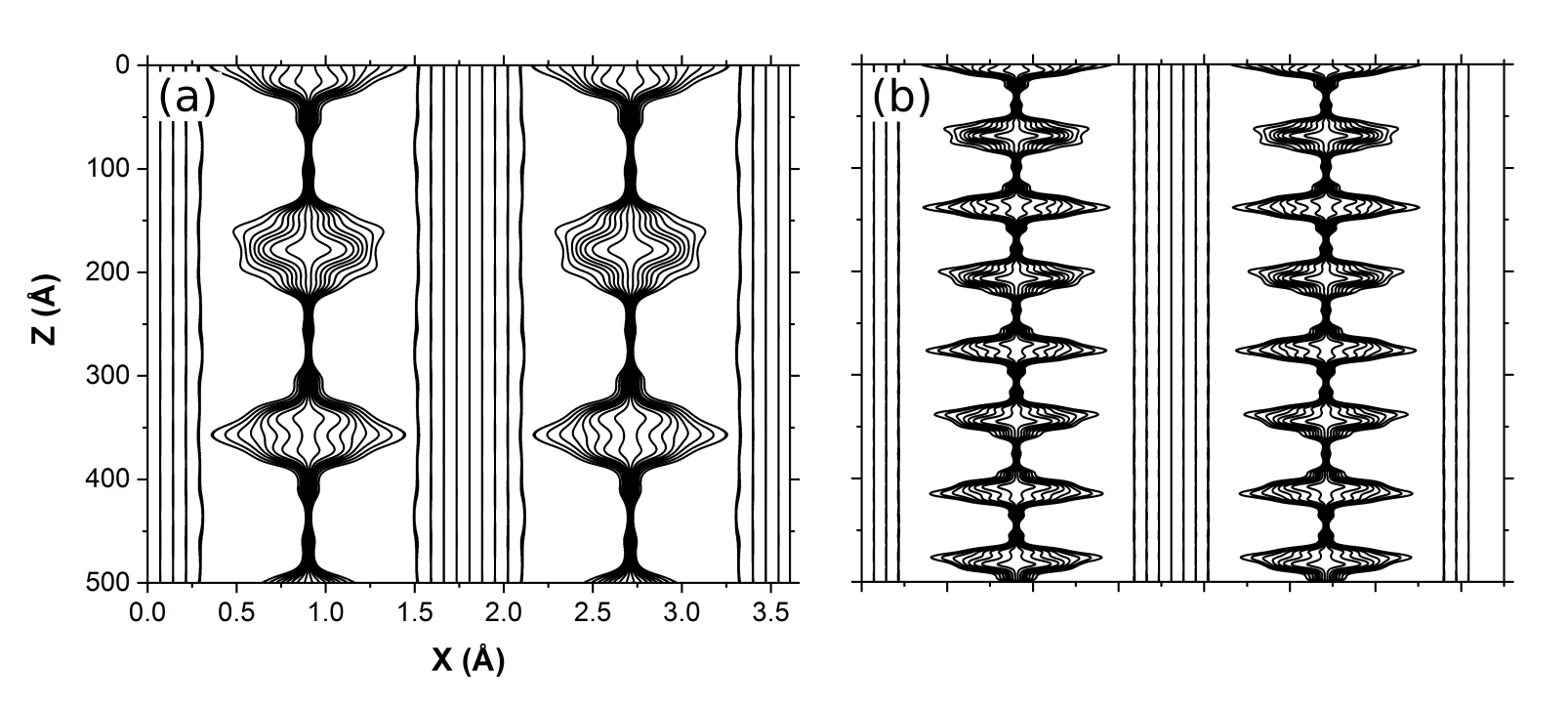}
 \caption{Trajectories propagated in the $\vg=(100)$ systematic row condition
 with $k_t=0$ at electron energies of (a) 200 keV and (b) 30 keV.}
 \label{fig:SRtraj}
\end{figure*}

\begin{figure*}
 \includegraphics[width=\textwidth]{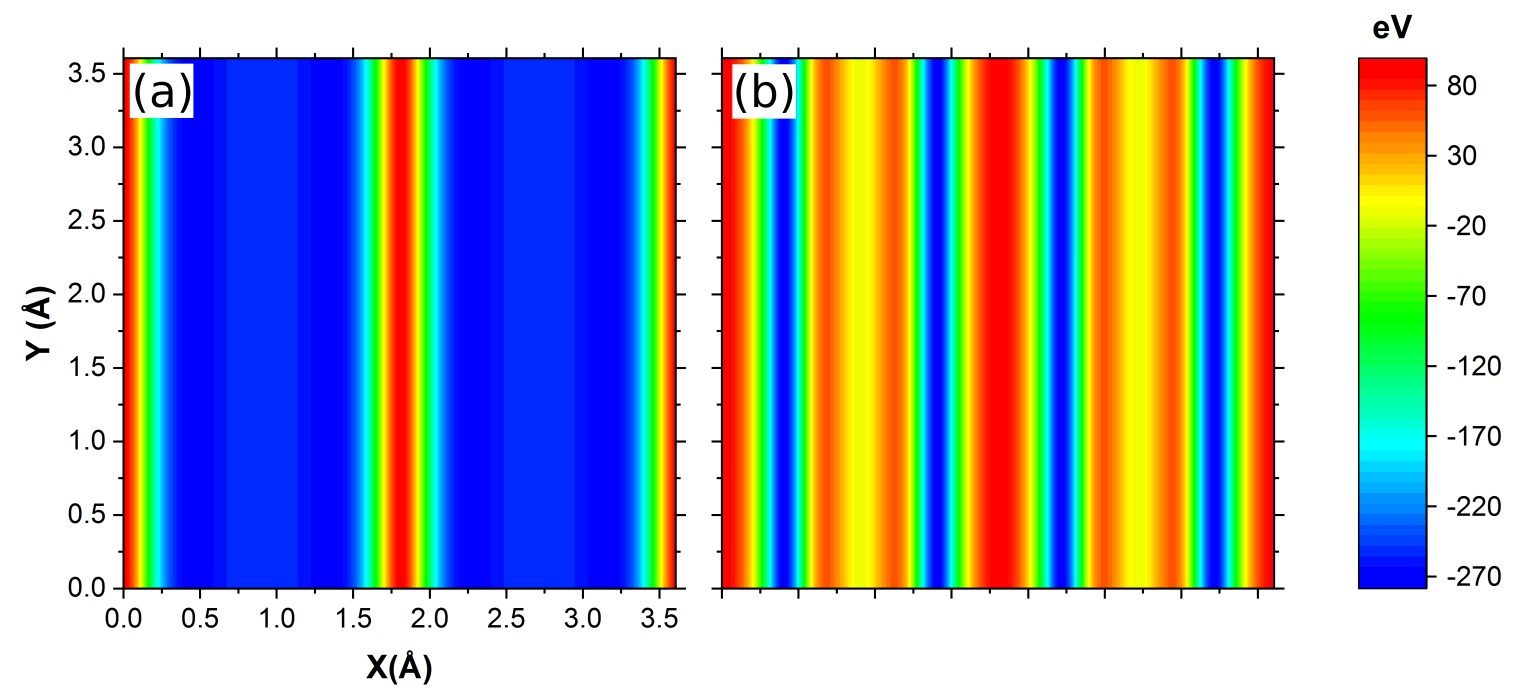}
 \caption{(a) Intensity distribution of exit wave function at 500 \AA\ in the
 $\vg=(100)$ systematic row orientation and (b) the corresponding quantum
 potential.} \label{fig:200keV-SR-EWQP}
\end{figure*}

 The special condition of the systematic row is
 important for convergent beam electron diffraction (CBED) patterns, simulations of
 bend contours, and defect analysis
 \cite{PhilipsPhylosophicalMag2011,RezPhylMagA1977,de2003introduction}. When
 multiples of a beam $\vg$ are excited, a material is said to be in a
 systematic row orientation \cite{de2003introduction}. Contrasts such as those
 found in bend contours are generated by the excitation of higher orders of a
 single beam where multiple such reflections are in the Bragg condition
 \cite{de2003introduction}. In CBED imaging, the contrasts within a single spot
 present themselves as bands. This is
simulated by including only multiples of a single beam direction in the dynamic
calculation \cite{de2003introduction}. If the simulation is done over
increments of the transverse wave component such that $\vk_t/\vg$ is
incremented uniformly, then rocking curves may be generated
\cite{reimer2008transmission}. The interest here is of the interaction of the traveling
particles through a material when they are diffracted in such a condition. The $\vg=(100)$
systematic row was simulated at normal incidence where 7 beams were considered,
from $-3\vg$ to $3\vg$. Figure \ref{fig:SRtraj} displays trajectories obtained
at both 200 keV and 30 keV.  
The trajectories along the atom columns are propagated straight through with
little deviation while those between the atom columns deviate towards the
columns and back towards their center, replicating a similar case to that of
normal incidence. Again, initial positions were chosen in a systematic way to
map out the entire beam. In reality, the uncertainty of the
initial position would dictate whether the electron remains in the band or
deviates between them. The exit wave function and quantum potential of the
simulation performed at 200 keV are displayed in Figure
\ref{fig:200keV-SR-EWQP}. The differences in intensities from the center of the
bands outwards are caused by the contributions of electrons passing between the
columns. This is due to the effects of the quantum potential seen in Figure
\ref{fig:200keV-SR-EWQP} (b). The quantum potential is positive exactly in the
center of the bands visible in the exit wave function and decreases to a sink
as it approaches them. This creates a force that is constantly deviated to and
from the contract bands, creating the trajectory paths of Figure
\ref{fig:SRtraj}.  
The paths taken by the trajectories demonstrate the dynamic effects that would
not otherwise be generated using a kinematic theory and provide an explanation
for differences in contrast across such imaging conditions.

\section{Conclusions}
The quantum trajectory method was coupled with a Bloch wave calculation of the
transmitted wave function of an electron beam through a thin copper foil.
Simulations were performed in the zone axis case, the two beam condition and
the systematic row condition. It was shown that quantum trajectories can
provide useful insight into electron-matter interactions by displaying where
the particles may pass as they are transmitted through an imaged material. In
the zone axis orientation, it is shown that the electrons are channeled by the
constant attractive and repulsive force exerted by the quantum potential that
surrounds the atom columns. In the two beam condition, diffraction of the plane
wave occurred by the quick separation of trajectories in to groups
corresponding to the diffracted and primary beams. Verifications of the two beam
calculations were made through mapping of the intensities of each the primary
beam and the diffracted one, where the contributions of each were shown to
oscillate continuously. Finally, in the case of the systematic row, the
contrast around the bands was explained by
oscillations in the quantum trajectories between atom columns. In total, the
method of associating trajectories to the propagating wave function was shown
to describe particle scattering with the involvement of quantum effects and can
be coupled in the future to Monte Carlo techniques to create all encompassing
image and particle simulations in electron microscopy.

\section*{Acknowledgments}
 The authors would like to thank Marc DeGraef and Scott Findlay for their insightful and helpful discussions.

\bibliography{bibliography}
%%%%%%%%%%%% The bibliography %%%%%%%%%%%%%%%%%%%%%

\end{document}